\documentclass{ws-p8-50x6-00}
\def\Pom{{\bf I\!P}}

\begin{document}

\title{ Possible Odderon discovery at HERA via 
charge asymmetry  in the diffractive $\pi^+\pi^-$ production}

\author{\underline{I.P.~Ivanov}, N.N.~Nikolaev}

\address{Institut f. Kernphysik,
Forschungszentrum J\"ulich, D-52425, J\"ulich, Germany\\ 
E-mail: i.ivanov@fz-juelich.de; n.nikolaev@fz-juelich.de}

\author{I.F.~Ginzburg}

\address{Sobolev Institute of Mathematics,
SB RAN, 630090, Novosibirsk, Russia\\
E-mail: ginzburg@math.nsc.ru}


\maketitle

\abstracts{We discuss how the evasive Odderon signal can be enhanced 
by final state interactions. We suggest the charge asymmetry of pion 
spectra in diffractive $\pi^{+}\pi^{-}$ photoproduction as a promising
signature of the Odderon exchange.}

\section{Introduction}

The Odderon --- a $C$-odd Regge singularity 
with intercept close to 1 ---
is an inevitable feature of QCD motivated picture of
high energy scattering and its experimental discovery and
the determination of its coupling to hadrons is crucial for 
the QCD theory of strong interactions. Within perturbative QCD, 
the pomeron exchange is naturally modelled by
the color-singlet two-gluon exchange in the $t$-channel.

Although much effort has been put forth to
verify the Odderon existence, up to now the Odderon remains elusive.
After years of vain search for the Odderon signal 
in particle-antiparticle cross section differences, 
the focus of investigation shifted towards processes initiated
exclusively by Odderon, namely, photoproduction of scalar or tensor
mesons $\gamma p \to \{\pi^0,\,\eta,\,\eta_c\,f_2\}+p'$.
However the cross sections involved are quite low and the signal
seems hardly separable from the background.

\section{The idea}

We propose a novel idea, inspired partly by earlier work \cite{brodsky} 
of how to catch a glimpse of Odderon existence.
We suggest to look at dipion, not any concrete meson, photoproduction
process $\gamma p \to \pi^+\pi^- p'$ at high energies. Then, the Odderon
will reveal itself as {\em the only source of kinematical charge asymmetries
in the dipion spectrum}.

Indeed, such a reaction can proceed via two types of channels.
The first type involves the Pomeron exchange, which helps
the initial photon turn into vector or higher spin mesons
with $C=-1$, which then decay into $\pi^+\pi^-$ pair.
The second type has Odderon as the $t$-channel exchange,
which lets the photon transform into scalar or tensor mesons with $C=+1$
with their subsequent decay into dipion.

Obviously, the two mechanisms of dipion production do not interfere
in the total cross section, but the interference effect
is still present in the differential cross sections:
$d\sigma \propto |A_{\Pom}|^2 + 2 Re(A_{\Pom}A_{O}^*) + |A_O|^2$. 
So, the essense of our idea is to extract this interference
term.

However, if we try to tackle this problem at the partonic level 
(that is thinking that we produce $q\bar q$ pair by either pomeron or Odderon
without asking how quarks will hadronize), we immediately run into
a difficulty. Indeed, at the partonic level
the pomeron induced amplitude is predominantly imaginary, while the Odderon
induced amplitude is mostly real. This makes the interference
term suppressed \cite{brodsky} by small factor 
$\sin\left[\pi(\alpha_\Pom-\alpha_O)/2\right]$, which is in addition
sensitive to the yet unknown value of Odderon intercept.

A closer look, however, shows that this is a fake problem.
Indeed, in reality we produce mesons, hadronic resonances, but not quarks. 
Introduction of intermediate resonances instantaneously
brings extra phases --- the Breit-Wigner phases --- and
removes the spurious orthogonality of the pomeron and Odderon amplitudes.

This observation can be viewed as an example of {\em strong breaking 
of parton-hadron duality}: the final-state interaction
has such a profound effect, that is gives birth to a new effect. 

In order to extract charge asymmetric terms in differential cross sections,
we suggest to perform their Fourier analysis.
In doing so, we find it particularly useful to introduce
two types of asymmetries: the forward-backward (FB) 
and the transverse ($\perp$) asymmetries defined according to:
$ A_{FB} = \int d\sigma \cos\theta / \int d\sigma\,; \quad
 A_{\perp} = \int d\sigma \cos\delta / \int d\sigma$, where $\theta$
and $\delta$ are the polar and the azimuthal angles of $\pi^+$ in
dipion rest frame. The physical meaning of
non-zero $A_{FB}$ is that one of pions tends to move
faster than the other, while $A_\perp$ would mean that
one of pions is produced preferrably along, or opposite to,
momentum transfer $\Delta_\perp$.

\section{A promising case: $\rho/f_2(1270)$ interference}

The above paragraphs contain already the key idea of our proposal.
A numerical illustration can give a feeling of the magnitude of the effect.

As said above, the production of the dipion can proceed via
formation of a whole number of intermediate resonances.
However it is known that the whole $(\pi^+\pi^-)_{C=-1}$ spectrum
around $M\sim 1$ GeV is overwhelmed by the $\rho$ peak, while
the most prominent feature of the $(\pi^+\pi^-)_{C=+1}$ spectrum
in this region is the $f_2(1270)$ resonance. Therefore it seems
quite a reasonable approximation to consider only two diagrams:
those with Pomeron induced $\gamma \to \rho$ transition
and with Odderon induced $\gamma \to f_2(1270)$ transition.

Since the helicity properties of the Odderon exchange are
not yet established, we considered two extreme cases:
(1) strict $s$-channel helicity conservation (SCHC)
for both $\rho$ and $f_2$, and (2) maximal $s$-channel helicity 
non-conservation (SCHNC) when $f_2$ meson is produced only 
in $\pm 2$ helicity state. In the former and the latter cases 
we expect FB and transverse asymmetries respectively.

Within the framework of SCHC, 
the two properly normalized amplitudes, written explicitly in terms of 
corresponding cross sections, are
\begin{eqnarray}
{\cal A}_{\Pom} &=&
\sqrt {\sigma_{\rho}}\ \mbox{e}^{{i\pi \over 2}\alpha_\Pom}
{ \sqrt{m_\rho \Gamma_\rho/\pi} \over M^2 - m_\rho^2 + i
m_\rho\Gamma_\rho}\cdot
 \sqrt{3} \sin\theta \cos(\widehat{\vec{e}_\perp\vec{r}_\perp}) 
\sqrt{B_{\rho}}\ \mbox{e}^{-{B_{\rho}\vec{\Delta}_\perp^2 \over 2}}\,,
\nonumber\\
{\cal A}_{\cal O} &=&\sqrt{\sigma_{f}} 
\ \mbox{e}^{i{\pi\over 2}\alpha_{\cal O}} 
{ i \sqrt{m_f \Gamma_f BR/\pi } \over  M^2 - m_f^2 + i m_f\Gamma_f }
\cdot \sqrt{15} \sin\theta \cos\theta 
\cos(\widehat{\vec{e}_\perp\vec{r}_\perp})
\sqrt{B_{f}}\ \mbox{e}^{-{B_{f}\vec{\Delta}_\perp^2 \over 2}}\,.\nonumber
\end{eqnarray}
One sees that Breit-Wigner phases present here
insure maximal interference effects directly under
$\rho$ peak (where both amplitudes become almost real)
and directly under $f_2$ peak (where both amplitudes turn predominantly
imaginary). In the former case, however, the charge symmetric contribution
is huge, which greatly suppresses the local FB asymmetry defined as
$A_{FB}(M^2) = \int d\sigma(M^2)\cos\theta /
\int d\sigma(M^2)$. On the contrary, under $f_2$ peak
the charge symmetric contribution is still saturated by the disappearing
$\rho$ meson tail, and therefore one can expect the largest
$A_{FB}(M^2)$ here. The value of this FB asymmetry
directly under $f_2$ resonance can be then estimated as
$$
A_{FB}(m_f^2) \approx \sqrt{{2 \over 5}} \cdot \sqrt{
{ \sigma_{f} \over \sigma_\rho}}
{m_f^2 - m_\rho^2 \over \sqrt{m_\rho m_{f} \Gamma_\rho \Gamma_{f} }}
\sqrt{BR(f_2\to \pi^+\pi^-) }.
$$
Here for simplicity we took $B_\rho=B_{f_2}$.
Plugging in numbers and using $\sigma_\rho = 6\mu b$ 
and $\sigma_O=20$ nb \cite{berger}, we
predict the peak value of local FB asymmetry as high as
$A_{FB}(m_f^2) \approx 21\%$. If one asks for the integrated
asymmetry in the optimal region $1.1 \mbox{ GeV} < M < 1.4 \mbox{ GeV}$,
one gets $ A_{FB} = 11\%\,.$

We can also analyze the case of maximal SCHNC, i.e. when 
$f_2$ meson is produced only in $\pm 2$ helicity state
(as indeed indicated by some model calculations \cite{berger}).
Cross section $\sigma_f \approx 20$ nb is now ascribed to this production.
The amplitudes ${\cal A}_{\Pom}(\rho^{+1})$ and ${\cal A}_{\cal O}(f_2^{+2})$
can be written in an analogous form.
Local transverse asymmetry at the $f_2$ peak is 
found to be $A_{\perp}(m_f^2)\approx 41\%$.
The integral transverse asymmetry within the invariant mass range 
$1.1 \mbox{ GeV} < M < 1.4 \mbox{ GeV}$ adds up to
$A_{\perp} = 21\%$. 

\section{Competitive mechanisms}

In order to safely interprete the charge asymmetry as 
the manifestation of the Odderon, we must make sure that
no other competitive mechanism can lead to the same effect,
at least with that large magnitude.

There are two contenders: the secondary reggeon ($\rho/\omega$)
and the photon exchanges. However both contributions are suppressed
so that  in the HERA energy domain both $\sigma^{\rho/\omega}(\gamma p 
\to f_2 p)$ and $\sigma^{\gamma}(\gamma p \to f_2 p)$ lie below 1 nb.
Thus we arrive at the conclusion:
{\em if the Odderon-induced cross section is as strong as expected
($\sigma^O(\gamma p \to f_2 X) \sim 20\mbox{ nb}$), then
 no other mechanism can mimic the Odderon signal}. 

\section{Conclusions}

In this work we suggested a radically different way
to search for much needed but still elusive Odderon:
that is, to look at charge asymmetries in the spectrum
of photoproduced dipions.

We also observed in our particular problem remarkably
severe breaking of much-celebrated parton-hadron duality.
The origin of this breaking was the strong final state interaction of
produced quark-antiquark pair.

Our main finding can be states as:
regardless of the nature of Odderon exchange, regardless
of the precise helicity structure of the Odderon coupling to matter,
irrespective of the exact value of Odderon intercept,
{\em the Odderon will be seen},
unless some tremendous unforeseen suppressions come into play. 
Borrowing the Odderon-induced
cross sections from models, we predict charge asymmetries to be
as huge as $10-20\%$. This effect is so prominent and so clear,
that it must be easily seen at HERA.

\end{document}